\documentstyle[12pt]{article}
\newcommand{\eq}[1]{eq.~(\ref{#1})}
\newcommand{\beq}{\begin{equation}}
\newcommand{\eeq}{\end{equation}}

\newcommand{\la}[1]{\label{#1}}

\newcommand {\bear}{\begin{eqnarray}}
\newcommand {\ear}{\end{eqnarray}}
\newcommand {\bds}{\begin{description}}
\newcommand {\eds}{\end{description}}
\newcommand {\riar}{\rightarrow}

\newcommand {\Go}{V_1(J)}
\newcommand {\Eo}{V_0(J)}

\newcommand {\FRG}[1]{\sum \frac{\Go}{M_1^{{#1}}}}
\newcommand {\FRO}[1]{\sum \frac{\Eo}{M_0^{{#1}}}}

\begin{document}
\begin{center}
{\Large Contribution of the Sixth
 Order Effective Chiral Lagrangian to the $\pi K$
 Scattering at Large $N_c$}\\
\vspace{0.95cm}
A.A.~Bolokhov$^{(1)}$, A.N.~Manashov$^{(1)}$, M.V.~Polyakov$^{(2)}$ and
 V.V.~Vereshagin$^{(1)}$ \\
\vspace{0.95cm}
{\it (1) Institute of Physics, Sankt-Petersburg University,
 198904 St. Petersburg, Russia \\
(2) Theory Division of Petersburg Nuclear Physics Institute,
188350 Gatchina, Russia}
\end{center}
\begin{abstract}
Using the method of asymptotic sum rules we estimated
the size of $O(m_s p_\pi^4)$  and $O(m_s^2 p_\pi^2)$ corrections to $\pi K$
scattering amplitude in large $N_c$ limit.  These corrections arise from
 the sixth order effective chiral lagrangian (EChL).  Our method enables us
 to estimate the corresponding terms of the sixth order EChL   in leading
 order of $1/N_c$ expansion in model independent way.  We found that the
corrections numerically are suppressed in spite of naive expectation of
30--35\%. Our estimation gives  the value of these corrections about
5--10\% .

\end{abstract}

\newpage
\noindent
{\bf 1.}
The technique of the Effective Chiral Lagrangians (EChL) provides us with
systematic way of low-energy expansion of correlators of different
colourless currents in the QCD \cite{Wei,GL}. The information about large
distance behaviour of the Quantum Chromodynamics is hidden in a set
of coupling constants which is finite if we restrict ourselves by finite
order  in momentum expansion.

In the lowest order of momentum expansion $O(p^2)$ the interactions
of (pseudo)goldstone mesons (pions, kaons and eta mesons) are
described by the famous Weinberg lagrangian \cite{Wei,WeiOld}:
\beq {\cal L}^{(2)}=\frac{F_0^2}{4}tr(\partial_\mu U^\dagger
\partial_\mu U) + \frac{F_0^2 B_0}{4} tr(\chi), \la{ecl2} \eeq
where $\chi=2 B_0 (\hat{m}U+U^\dagger\hat{m} )$,
$\hat{m}=diag(0,0,m_s) $ is a quark mass matrix and
$F_0$ and $B_0$ are low-energy coupling constants contained
an information about long-distance behaviour of the QCD.
The latter can be either extracted from an experiment or
calculated in the framework of some models of strong interactions.

In the next $O(p^4)$ order the interactions of the (pseudo)goldstone
mesons are described by the following EChL \cite{VVV,GLsu3}
\[ {\cal
L}^{(4)}=L_1 \left[ tr(\partial_\mu U^\dagger \partial_\mu U)\right]^2 +
 L_2
tr(\partial_\mu U^\dagger \partial_\nu U)\cdot tr(\partial_\mu U^\dagger
 \partial_\nu U) + \]
\[+
 L_3  tr(\partial_\mu U^\dagger \partial_\nu U
\partial_\nu U^\dagger \partial_\mu U)
+L_4 tr(\partial_\mu U^\dagger \partial_\mu U)\cdot tr( \chi) + \]
\beq
+L_5 tr(\partial_\mu U^\dagger \partial_\mu U \chi)
+L_6tr(\chi)^2
+L_7tr(\hat{m} U^\dagger + U \hat{m}^\dagger)^2
+L_8tr(\chi^2)
+L_8tr(\chi^2).
\la{ecl4}
\eeq
Here new eight coupling constants $L_{1 \dots 8}$ are appeared.
These coupling constants enter the mass splitting in the
(pseudo)goldstone octet, $d$-wave $\pi \pi$ and $\pi K$
scattering, etc. They have been calculated
by integration of non-topological axial anomaly in QCD
\cite{DiaEid,Bal,And} and in the instanton liquid model
of QCD vacuum \cite{DiaPet}.
Throughout the paper we shall neglect  the masses of light
$u $ and $d$ quarks and we shall work in the leading
order of $1/N_c$ expansion, so we can put
 $L_4=L_6=2L_1-L_2=0$ \cite{GLsu3}.
Hence in this order  the (pseudo)goldstone meson interactions in
the fourth order are described by six universal parameters of the EChL--
$F_0$ , $B_0$, $L_2$, $L_3$, $L_5$ and $L_8$.

Using the EChL eqs.(\ref{ecl2}) and  (\ref{ecl4}) one can calculate, say,
$\pi K$ scattering amplitude to the orders
$O(p_\pi^4)$, $O(m_s p_\pi^2)$ and $O(m_s^2)$,
 where $p_\pi$ is pion momentum.
 The corrections of the form $O(m_s p_\pi^4)$, $O(m_s^2 p_\pi^2)$ and
$O(m_s^3)$ due to
higher order EChL naively are expected to be
of order $O(m_K^2/m_\rho^2)$ i.e. about 30--35\%.
To calculate the size of these corrections one needs to know the sixth
 order EChL. Unfortunately the knowledge of the higher order EChL is
rather poor.
In this paper we calculate the size of these corrections to
$\pi K$ scattering amplitude of the orders $O(p_\pi^4)$
and $O(m_s p_\pi^2)$,
 using
the  method of asymptotic sum rules \cite{p5}.

Method of asymptotic sum rules for
the (pseudo) Goldstone elastic scattering in large $N_c$ limit
was suggested in \cite{p5}. In a sense, it is based on the
generalization of Weinberg's asymptotic restrictions for
chiral amplitudes \cite{p6} for the case of nonzero (pseudo)
Goldstone mass. This method allows to express the
parameters of the effective chiral lagrangian through the
masses and widths of the admissible resonances and put a
dynamical restrictions on resonance spectrum at large $N_c$
\cite{p5}. We apply this method to
particular case of $\pi \pi$ and $\pi K$ elastic processes.

Using this method
we find that the  corrections to $\pi K$
scattering amplitudes  of the form $O(m_s p_\pi^4)$ and $O(m_s^2 p_\pi^2)$
($p_\pi$ is soft pion momentum)
arising from the sixth order EChL
numerically are suppressed in spite of
naive expectation of 30--35\%. Our estimation gives  the value of
these corrections about 5--10\%, so one can use EChL (\ref{ecl4})
for, say, calculation of d--wave $\pi K$ scattering amplitude with
good accuracy.

\vskip 0.3cm
\noindent
{\bf 2.}
Let us consider the elastic $\pi \pi$--scattering process
$$
\pi_a(k_1)+\pi_b(k_2)\rightarrow\pi_c(k_3)+\pi_d(k_4).
$$
( $a,b,c,d=1,2,3$ are the isotopic indices and
$k_1,..,k_4$ --- pion momenta.)
Its amplitude $ M^{abcd} $  can be written in the form:
\beq
M^{abcd}=\delta^{ab}\delta^{cd}A+\delta^{ac}\delta^{bd}B+
\delta^{ad}\delta^{bc}C ,
\label{1}
\eeq
where $A,B,C$ are the scalar functions of Mandelstam variables
 $s,t,u$:
\beq
s=(k_1+k_2)^2,\qquad  t=(k_1-k_3)^2,\qquad  u=(k_1-k_4)^2,
\la{mv} \eeq
obeying the Bose--symmetry requirements:
\begin{eqnarray}
A(s,t,u) &=& A(s,u,t) \ , \nonumber \\
B(s,t,u) &=& A(t,s,u) \ , \label{2} \\
C(s,t,u) &=& A(u,t,s) \ . \nonumber
\end{eqnarray}
The amplitude of the $\pi K$ scattering process
\beq
\pi_a(k_1) +K_\alpha(k_2) \rightarrow \pi_b(k_3) +K_\beta(k_4).
\eeq
can be expressed in terms of two (iso)scalar functions $A_{+}(\nu,t) $
and $A_{-}(\nu,t)$ by
\beq
M^{ab}_{\alpha \beta}=\delta^{ab} \delta_{\alpha \beta} A_{+}(\nu,t)+
i \epsilon^{abc} \sigma^c_{\alpha \beta}A_{-}(\nu,t),
\eeq
where invariant variable $\nu=s-u$ is expressed via Mandelstam variables
\eq{mv}.
Near threshold of the reactions one can expand the (iso)scalar
amplitudes $A(s,t)$ ,  $A_{+}(\nu,t)$ and $A_{-}(\nu,t)$
in power series of pion momenta:
\beq
 A(s,t)=\sum_{i,j}^{} a_{ij} s^i t^j , \label{19}
\eeq
\beq
A_{+}(\nu,t)=a_0(m_s) + a_1(m_s)\cdot t + a_2(m_s)\cdot t^2
+ a_3(m_s)\cdot \nu^2 t+
\dots,
\eeq
\beq
A_{-}(\nu,t)=b_1(m_s)\cdot \nu + b_2(m_s)\cdot \nu t +
\dots
\eeq
Non-analitic parts of the amplitudes (like $E^4 log(E)$) are
suppressed by additional $1/N_c$. Parameters of the near threshold
expansion of the
$\pi K$ scattering amplitude depend on strange quark mass.

{}From the Effective Chiral Lagrangian \eq{ecl2},\eq{ecl4} one gets
an expression for the low-energy parameters of the $\pi \pi $
scattering amplitude
and for expansion in $m_s$ of the corresponding parameters of the
$\pi K$ scattering amplitudes in terms of universal parameters
of the fourth order EChL-- $F_0$ , $B_0$, $L_2$ and $L_3$ \footnote{
$L_5$, $L_7$ and $L_8$ do not enter explicitely into these quantities.
Dependents on them apears in one--loop order \cite{Bernard} i.e. in
next order of the $1/N_c$ expansion}:
$$
a_{00}=0; \>\>\>\> a_{10}=\frac{1}{F_0^2}; \>\>\>\> a_{01}=0;
$$
$$
a_{20}=\frac{4(2L_2+L_3)}{F_0^4}; \>\>\>\>\>\>\>
a_{11}=a_{02}=\frac{8L_2}{F_0^4} \ , $$
\beq
b_1(m_s)=-\frac{1}{F_0^2},
\eeq
(It easy to prove that the corrections due to the non-zero
strange quark mass to $b_1$ should vanish by chiral Ward identity
 i.e. $\frac{d}{dm_s} b_1(m_s) \equiv 0 $)
\beq
a_3(m_s)=\frac{8(L_3+4L_2)}{F_0^4}\cdot(1+m_s \xi_1) +O(m_s^2).
\eeq
Here we introduce the quantity $\xi_1$ which determines
the values of $m_s$-corrections to the low-energy parameters of
$\pi K$ amplitude of order $O(p^4)$. This quantity can be calculated
in terms of parameters of the sixth order EChL, unfortunately the
EChL to these order is not known.
In next section, using method of asymptotic sum rules \cite{p5},
we shall express these quantities via
parameters (masses and widths) of $\pi \pi$ and $\pi K$ resonances and show
that $m_s \xi_1 \sim 0.1$ what justified an applicability of $m_s$-expansion
for the $\pi K$ amplitude of order $O(p^4)$
 (applicability of the ECL (\ref{ecl2},\ref{ecl4}) for description of
 low--energy $\pi K$ elastic scattering with an accuracy of 10\%).

\vskip 0.3cm
\noindent
{\bf 3.}
In this section we derive sum rules (SR) relating low-energy coeffcients
$a_{ij}$ (for $\pi\pi$ amplitude), $a_i(m_s)$ and $b_i(m_s)$
(for $\pi K$ amplitude) to parameters (masses and widths) of
 $\pi\pi$ and $\pi K$ resonances in the leading order    of the $1/N_c$
expansion.
To do this we use the Weinberg's method of the asymptotic restrictions
\cite{p6} generelized for the case of nonzero mass of (pseudo)goldstone
bosons  and for the case of scattering on nonzero angle \cite{p5}.
 Idea of the method is very simple--
  in the large $N_c$ limit we can represent $\pi K$ and $\pi\pi$
elastic scattering amplitudes as an infinite sum of resonance pole
contributions, in this limit the amplitude is free of other
singularities.  Then one assumes that the asymptotic behaviour of the
amplitude at $s \rightarrow \infty$ and fixed $t$ is not worse than the
one in Regge's theory. This assumption should be fulfilled in the large
$N_c$ limit to ensure the renormalizability of the underlying theory - QCD.
Using the resonance form of the amplitude one can express coefficients of
near threshold expansion of the amplitudes through the masses and widths of
 $\pi\pi$ and $\pi K$ resonances.
\bear
& & 1/F_0^2 = \FRG{4}+\FRO{4}  \ ; \label{pipiSR1}   \\
& & 4L_2/F_0^4= \FRG{6} \ ;  \\
& & 4(2L_2+L_3)/F_0^4 = -\FRG{6}+\FRO{6}
\ , \label{pipiSR2}
\ear
where $M_I(0)$  ($I$=1,2) are masses of the $\pi\pi$ resonances with
 isospin $I$ and constants
$V_I(J)$ can be
related to the decay widths of corresponding resonances with
 isospin $I$ and spin $J$:
$$
V_1(J)=8\pi(2J+1)\cdot \frac{M_1^2}{k_{\pi}}\Gamma(R \riar \pi
\pi) \ ,
$$
$$
  V_0(J)=\frac23 \cdot 8\pi(2J+1)\cdot
\frac{M_0^2}{k_{\pi}}\Gamma(R \riar \pi \pi) \ ,
$$
$k_{\pi}$ being the pion CMS momentum.
\beq
b_1(m_s)=\frac{1}{F_0^2}=\sum \frac{4R(m_s)}{(M_R(m_s)^2-m_K^2)^2},
\la{piKSR1}
\eeq
\beq
a_3(m_s)=\frac{8(L_3+4L_2)}{F_0^4}\cdot(1+m_s \xi_1) +O(m_s^2) =
\sum \frac{8R(m_s)}{(M_R(m_s)^2-m_K^2)^3},
\la{piKSR2}
\eeq
$R(m_s)$ and $M_R(m_s)$ are residues and masses of the $\pi K$-resonances.
\beq
R(m_s)=\frac13 8\pi(2j+1)\cdot \frac{M_R}{k}\Gamma(R \rightarrow \pi K),
\eeq
\beq
k=\frac{1}{2M_R}\sqrt{(M_R^2-(m_K+m_\pi)^2)(M_R^2-(m_K-m_\pi)^2 )}.
\eeq
When $m_s \rightarrow 0$ one gets:
\beq
R(m_s)=\frac14 V_I\cdot(1+m_s \Delta +O(m_s^2)),
\eeq
(here $I=0$ for even spin resonances, $I=1$ for odd spin resonances)
\beq
M_R(m_s)^2=M_R(0)^2 \cdot(1+m_s \delta +O(m_s^2)).
\eeq
The quantities $\delta$ and $\Delta$ are related to the mass and
widths splitting inside resonance nonet.
For example, for well established $\rho$-meson octet one has:
\beq
m_s \Delta \approx
 \frac{4R(K^* \rightarrow \pi K)-
V_1(\rho \rightarrow \pi\pi)}{V_1(\rho \rightarrow \pi\pi) }
 = -0.25,
 \eeq
\beq
m_s \delta \approx \frac{M_{K^*}^2-M_\rho^2}{M_\rho^2}=0.35.
\eeq

Expanding the eqs.(\ref{piKSR1},\ref{piKSR2}) in powers of $m_s$
one can express $m_s \xi_{1}$ via resonance parameters.
{}From \eq{piKSR1} one gets:
\beq
0=\frac{\partial b_1(m_s)}{\partial m_s}= \langle m_s \Delta-2\cdot(m_s
\delta-\frac{m_K^2}{M_R(0)^2}) \rangle_2,
\la{cor2}
 \eeq
where we introduce  notations
\beq
\langle F \langle_k \equiv \frac{\sum_{I=0,1} \sum_{res}
 \frac{V_I}{M_R(0)^{2k}} F }{\sum_{I=0,1}
 \sum_{res} \frac{V_I}{M_R(0)^{2k}} }
\eeq
for any quantity $F$ depending on resonance parameters.

Naively one can expect that relative corrections due to nonzero
strange quark mass to $b_1(m_s)$ must be of order
$m_s \delta \approx (M_{K^*}^2/M_\rho^2-1)$ (e.i. $\sim 0.35$
for realistic resonance spectrum) but the chiral Ward identities require
that (see \eq{piKSR1}) resonance spectrum must be adjusted to provide exact
   cancelation of different corrections in any order of $m_s$-expansion
   each of those is not small.
 Expanding rhs of the sum rules (\ref{piKSR1},\ref{piKSR2})
 in power of $m_s$
 one can express $\xi_{1}$ in terms of resonances parameters:
\beq
m_s \xi_1=<m_s \Delta-3\cdot(m_s
\delta-\frac{m_K^2}{M_R(0)^2})>_3.
\la{cor4}
 \eeq
In the previous (second) order we found complete cancellations among
different corrections of order $m_s$, so we could, in principle, expect
similar cancellations in considered (fourth) order.

To estimate numerically $m_s \xi_1$ we use two different methods.
In the first one we use the typical parameters of the $\rho$-meson
nonet to estimate $m_s \xi_1$ by \eq{cor4}.
The result is:
\[
m_s \xi_1 \approx
 \frac{4R(K^* \rightarrow \pi K)-
V_1(\rho \rightarrow \pi\pi)}{V_1(\rho \rightarrow \pi\pi) } -3\cdot
\frac{M_{K^*}^2-(M_\rho^2+m_K^2)}{M_\rho^2} =\]
\beq
=-0.25+3\cdot0.06=-0.07.
\label{prikid}
\eeq

In the second method we calculate $b_1(m_s)$ and $a_3(m_s)$ directly
by eqs.(\ref{piKSR1},\ref{piKSR2}) using the realistic resonance spectrum
\cite{PDG} (results are presented in Table 1 for $\pi \pi$ SR
and Table 2 for $\pi K$ SR)
) and $L_2$, $L_3$ and $F_0^2$
by eqs.(\ref{pipiSR1}-\ref{pipiSR2}), and then extract $\xi_1$ using:
\beq
m_s\cdot \xi_1 \approx \frac{a_3(m_s)
\cdot F_0^4-8(L_3+4L_2)}{8(L_3+4L_2)}.
\eeq
{}From the Tables 1. and 2.  one can conclude that $m_s\cdot \xi_1=-0.085\pm
0.005$
what confirm our previous estimation (\ref{prikid}).
 Also we see  that the parameter
$b_1(m_s)$ does not depend on $m_s$ with good accuracy,
what is required by chiral Ward identities.
 Numerically both $\pi \pi$ and $\pi K$ sum rules (\ref{pipiSR2},
\ref{piKSR1}) give
$F_0^{(exp)}/F_0^{(SR)} =0.83\pm 0.1$. From the latter number one can
conclude that next to leading $1/N_c$ corrections to $F_0$ have
a size around
30\% and negative sign.\footnote{ An attempt to explain the observed
difference by missing resonance(s) is not likehood because in this
case one has to add resonance simultaneously in $\pi \pi$ and
$\pi K$ channels and hence we would have a light strange resonance
what seems to be excluded by data.}

\vskip 0.3cm
\noindent
{\bf 4.}
The applicability of the fourth order effective chiral lagrangian
(\ref{ecl2},{\ref{ecl4}) for the calculation
$\pi K$
scattering amplitudes up to order $O(p_\pi^4)$
with 5-10\% accuracy is checked at least
in large $N_c$ limit.
The $\pi K$
scattering amplitude calculated with use of fourth order EChL
has a relative corrections of order $m_s$ arising
from the sixth order EChL. Naively we would expect that the corrections
are of order $m_k^2/m_\rho^2 \sim 30-35\%$ (the same as for mass
splittings in $SU(3)$ meson multiplets), but, at least in the leading
order of $1/N_c$ expansion, these corrections are anomalously small.
Roughly speaking the actual behaviour of the corrections with $m_s$
is not $\sim m_k^2/m_\rho^2$, but $\sim
\frac{m_{K^*}^2-(m_\rho^2+m_K^2)}{m_\rho^2}$.
Though both these quantities are proportional to $m_s$ the latter has
additional
numerical suppresion.
Simultaneously we found that resonance spectrum has to satisfy an
equation (\ref{cor2}) to provide chiral Ward identities.
We checked that this equation is satisfied with quite good
accuracy on the known resonance spectrum \cite{PDG}.
We think our sum rules (\ref{pipiSR2},\ref{piKSR1},\ref{piKSR2})
for parameters of fourth order EChL and (\ref{cor2}) for resonance
spectrum are the good consistency check of different models of
QCD at large number of colours.

\newpage
\begin{table}
\begin{center}
\begin{tabular}{|c|c|c|c|}
\hline
 Resonance  & spin & $4 L_2 +L_3=\frac{a_3(m_s=0)*F_0^4}{8} $
&$ F_\pi^2/F_0^2 $  \\
\hline
$f_0(975)$& 0 &$ (1.98\pm 0.44)10^{-5}$  & 0.01\\
\hline
$f_0(1400)$ & 0 & $(3\pm2)10^{-5} $ & 0.03\\
\hline
$f_0(1590)$&0 & $(2.\pm2.)10^{-6} $ & $ 0.002$\\
\hline
$f_2(1275)$&2 & $(1.51\pm0.075)10^{-4} $ & $ 0.1$\\
\hline
$f_2(1525)^\prime$&2 & $\sim 0 $ & $ \sim 0.$\\
\hline
$f_2(1720)$&2 & $\sim 0 $ & $ 0.002$\\
\hline
$f_4(2050)$&4 & $\sim 0 $ & $ 0.001$\\
\hline
$f_6(2510)$&6 & $(1.2\pm0.6)10^{-6} $ & $ 0.001$\\
\hline
\hline
$\rho(770)$&1 & $1.66\cdot 10^{-3} $ & $ 0.51$\\
\hline
$\rho(1450)$&1 & $(1.2\pm 0.2) 10^{-4} $ & $ 0.04$\\
\hline
$\rho(1700)$&1 & $(2.\pm 3.) 10^{-6} $ & $ 0.003$\\
\hline
$\rho_3(1690)$&3 & $(2.5\pm 0.4) 10^{-6} $ & $ 0.03$\\
\hline
$\rho_5(2400)$&5 & $(2.7\pm 1.4) 10^{-6} $ & $ 0.003$\\
\hline
\hline
Sum &-& $(2.0\pm0.1)10^{-3}$& $\sim 0.7$\\
\hline
\end{tabular}
\end{center}
\caption{Contributions of known $\pi\pi$ resonances to sum rules
 eqs.(13,15)}
\end{table}

\newpage
\begin{table}
\begin{center}
\begin{tabular}{|c|c|c|c|}
\hline
 Resonance  & spin & $\frac{a_3(m_s)*F_0^4}{8}$&
$b_1(m_s) F_\pi^2=F_\pi^2/F_0^2 $  \\
\hline
$K^*(1430)$& 0 &$ (9.5\pm 2.0)10^{-5}$  & 0.08\\
\hline
$K^*(891)$ & 1 & $(1.53 \pm0.02)10^{-3} $ & 0.39\\
\hline
$K^*(1370)$&1 & $(1.\pm1.)10^{-5} $ & $ 0.01$\\
\hline
$K^*(1400)$&1 & $(1.\pm1.)10^{-6} $ & $ 0.001$\\
\hline
$K^*(1680)$&1 & $(7.\pm 6.)10^{-5} $ & $  0.086 $\\
\hline
$K^*(1430)$&2 & $(8.8\pm 0.9)10^{-5} $ & $  0.073 $\\
\hline
$K^*(1780)$&3 & $(2.2\pm 0.4)10^{-5} $ & $  0.03 $\\
\hline
$K^*(2045)$&4 & $(7.9\pm 2.3)10^{-6} $ & $  0.014 $\\
\hline
\hline
Sum &-& $(1.83 \pm 0.1)10^{-3}$& $\sim 0.7$\\
\hline
\end{tabular}
\end{center}
\caption{Contributions of known $\pi K$ resonances to sum rules
 eqs.(16,17)}
\end{table}
    \end{document}